\documentclass[12pt]{article}
\usepackage{graphics,epsfig}
\def\al{\alpha}

\def\ga{\gamma}

\def\Ga{\Gamma}

\newcommand{\beq}{\begin{equation}}
\newcommand{\eeq}{\end{equation}}
\newcommand{\bea}{\begin{eqnarray*}}
\newcommand{\eea}{\end{eqnarray*}}
\newcommand{\beaq}{\begin{eqnarray}}
\newcommand{\eeaq}{\end{eqnarray}}
\begin{document}

\centerline{\Large\bf Duality in $N=2$ Super-Liouville Theory}
\vskip 2cm
\centerline{\large Changrim Ahn\footnote{ahn@dante.ewha.ac.kr}$^{\dagger}$, 
Chanju Kim\footnote{cjkim@phya.snu.ac.kr}$^{\dagger}$, 
Chaiho Rim\footnote{rim@mail.chonbuk.ac.kr}$^{\ddagger}$, 
and M. Stanishkov\footnote{stanishkov@dante.ewha.ac.kr;
On leave of absence from Institute of Nuclear Research and Nuclear
Energy, Sofia, Bulgaria}$^{\dagger}$ }
\vskip 1cm
\centerline{\it$^{\dagger}$Department of Physics, Ewha Womans University}
\centerline{\it Seoul 120-750, Korea} \vskip .5cm
\centerline{\it$^{\ddagger}$ Department of Physics, Chonbuk National
University} \centerline{\it Chonju 561-756, Korea} \vskip 1cm
\centerline{\small PACS: 11.25.Hf, 11.55.Ds} \vskip 2cm
\centerline{\bf Abstract}
In this paper we consider a strong-weak coupling duality of 
the $N=2$ super-Liouville field theory (SLFT). 
Without the self-duality found in other Liouville theories,
the $N=2$ SLFT, we claim, is associated with a `dual' action
by a transformation $b\to 1/b$ where $b$ is the coupling constant.
To justify our conjecture, we compute the reflection amplitudes (or
two-point functions) of the (NS) and the (R) 
operators of the $N=2$ SLFT based on the conjectured dual action
and show that the results are consistent with known results.

\section{Introduction}
Two-dimensional Liouville field theory (LFT) appears naturally 
in the context of the 2D quantum gravity and string theories 
\cite{LFT,CurTho}.
This theory has been extended to the supersymmetric Liouville theories 
to accomodate world-sheet supersymmetries appearing in the string 
theories.
These Liouville-type theories are also interesting 
for quantum field theoretical properties as well.
They possess both conformal symmetry and the strong-weak coupling duality.
The strong-weak coupling duality has been attracting much attention
recently to understand non-perturbative aspects of the various quantum
field theories rigorously.
For example, the seminal work of Seiberg and Witten on 
the supersymmetric Yang-Mills theories in (3+1)-dimensions 
is based on the intuitive observation that the gauge theories in the
strong coupling limit are described by a weak coupling region of
some effective action \cite{SeiWit}.
More recently, similar duality arises in the string theory context 
to understand strings and branes nonperturbatively.
This duality also arises in statistical systems such as the 2D Ising model.

For the LFT and the $N=1$ SLFT, the strong-weak coupling duality appears
as a quantum symmetry closely connected to (super-)conformal symmetries.
It has been observed that the background charge is renormalized to
$Q=b+1/b$ by quantum corrections and the theories preserve the quantum
conformal symmetries \cite{CurTho,ORaf}.
This means the two LFTs are invariant under $b\to 1/b$, i.e. self-dual.
These two symmetries are essential to determine exact correlation functions
for the LFT \cite{ZamZam} and the $N=1$ SLFT \cite{RasSta}.
Now let us consider the duality symmetry of the $N=2$ SLFT.
CFTs with the $N=2$ supersymmetry have been actively studied mainly due
to possible applications to string theories.
In particular, the $N=2$ SLFT appears in the context of the black-hole 
solutions of a string theory \cite{n2slft}.
With or without the conformal symmetry, two-dimensional models with
$N=2$ supersymmetry show an interesting feature, namely,
the nonrenormalization.
The parameters in the supersymmetric action do not change in all orders
of perturbative calculations. 
This means that the $N=2$ SLFT maintains the conformal symmetry without
renormalization of the background charge and loses the self-duality.

Our main proposal in this paper is that 
the $N=2$ SLFT still shows an interesting duality behaviour.
Under the dual transformation $b\to 1/b$, the theory maps to 
a `dual' action which is another $N=2$ super CFT.
The $N=2$ SLFT with a strong coupling can be described by the dual action
perturbatively.
We compute the reflection amplitudes (the two-point functions) 
of the theory using functional relations derived from the actions.
This procedure provides an exact relation between two parameters
$\mu,{\tilde\mu}$, which fixes the relations between the two actions
completely.
To check a self-consistency of our proposal, we compare the reflection
amplitudes derived from the conjecture action with some independent
result derived in a totally different context.

\section{$N=2$ Super-Liouville theory and its Dual action}

The action of the $N=2$ SLFT at the flat background is given by
\beq
{\cal A}_{\rm I}(b)=\int d^2 z \left[\int d^4\theta SS^{\dagger}
+\mu\int d^2\theta e^{bS}+c.c.\right]
\label{actionone}
\eeq
where $S$ is a scalar superfield satisfying
\beq
D_{-}S={\overline D}_{-}S=0,\qquad D_{+}S^{\dagger}=
{\overline D}_{+}S^{\dagger}=0.
\eeq
The Lagrangian of the $N=2$ SLFT 
can be expressed in terms of the component fields as follows:
\begin{eqnarray}
{\cal L}&=&{1\over{4\pi}}\left[2\varphi\partial{\overline\partial}
\varphi^{\dagger}+2\varphi^{\dagger}\partial{\overline\partial}
\varphi+\psi^{\dagger}{\overline\partial}\psi+
\psi{\overline\partial}\psi^{\dagger}+
{\overline\psi}^{\dagger}{\partial}{\overline\psi}
+{\overline\psi}{\partial}{\overline\psi}^{\dagger}\right]\\
&-&{5\over{4}}\pi\mu^2 b^2 e^{b\varphi+b\varphi^{\dagger}}
+{\mu b^2\over{2}}\psi^{\dagger}{\overline\psi}^{\dagger}e^{b\varphi}
+{\mu b^2\over{2}}\psi{\overline\psi}e^{b\varphi^{\dagger}}.
\label{lagrangianone}
\end{eqnarray}
As in the LFT and the $N=1$ SLFTs, one should introduce a background charge 
$1/b$ so that the second term in Eq.(\ref{actionone}) becomes the screening
operator of the conformal field theory (CFT).
However, a fundamental difference arises where
the background charge is unrenormalized due to the $N=2$ supersymmetry.
For the LFT and the $N=1$ SLFTs, the background charge is 
renormalized to $Q=1/b+b$ and the theories are invariant under the 
dual transformation $b\to 1/b$. 
This self-duality plays an essential role to determine
various exact correlation functions of those Liouville theories. 
Unrenormalized, the $N=2$ SLFT is not self-dual.

This theory is a CFT with a central charge 
\beq
c=3+6/b^2.
\eeq
The primary operators of the $N=2$ SLFT are classified into the 
Neveu-Schwarz (NS) and the Ramond (R) sectors and can be written 
in terms of the component fields as follows \cite{Marian}:
\begin{equation}
N_{\alpha{\overline\alpha}}=
e^{\alpha\varphi^{\dagger}+{\overline\alpha}\varphi},\qquad
R^{\pm}_{\alpha{\overline\alpha}}=
\sigma^{\pm}e^{\alpha\varphi^{\dagger}+{\overline\alpha}\varphi},
\label{primary}
\end{equation}
where $\sigma^{\pm}$ is the spin operators.
The conformal dimensions of these fields are given by
\begin{equation}
\Delta^{N}_{\alpha{\overline\alpha}}=-\alpha{\overline\alpha}
+{1\over{2b}}(\alpha+{\overline\alpha}),\qquad
\Delta^{R}_{\alpha{\overline\alpha}}=\Delta^{N}_{\alpha{\overline\alpha}}+
{1\over{8}}.
\label{delta}
\end{equation}
The $U(1)$ charges are given by
\begin{equation}
Q^{N}_{\alpha{\overline\alpha}}=-{1\over{2b}}(\alpha-{\overline\alpha}),
\qquad
Q^{R\pm}_{\alpha{\overline\alpha}}=Q^{N}_{\alpha{\overline\alpha}}
\pm{1\over{4}}.
\label{u1charge}
\end{equation}
From these expressions, one can notice that 
\beq
\alpha\to 1/b-{\overline\alpha},\qquad
{\overline\alpha}\to 1/b-\alpha
\label{reflection}
\eeq
do not change the same conformal dimension and $U(1)$ charge. 
From the CFT point of view, this means that 
$N_{1/b-{\overline\alpha},1/b-\alpha}$ should be identified
with $N_{\alpha{\overline\alpha}}$ and similarly for the (R) operators
upto normalization factors.
The reflection amplitudes are determined by these normalization factors.

Without the self-duality, it is possible that there exists a `dual' action 
to (\ref{actionone}) whose perturbative (weak coupling) behaviours describe
the $N=2$ SLFT in the strong coupling region.  
This action should be another CFT.
Our proposal for the dual action is as follows:
\begin{equation}
{\cal A}_{\rm II}(b)=\int d^2 z \int d^4\theta \left[SS^{\dagger}
+{\tilde\mu}e^{b(S+S^{\dagger})}\right]
\label{actiontwo}
\end{equation}
with the background charge $b$.
The $N=2$ supersymmetry is preserved because $S+S^{\dagger}$ is a
$N=2$ scalar superfield.
One can see that this action is conformally invariant because the 
interaction term is a screening operator.
Our conjecture is that the two actions, 
${\cal A}_{\rm I}(b)$ and ${\cal A}_{\rm II}(1/b)$ are equivalent.
To justify this conjecture, we will compute the reflection amplitudes 
based on these actions and will compare with some independent results.

\section{The Reflection Amplitudes}

As mentioned above, the reflection amplitudes of the 
Liouville-type CFT is defined by the linear 
transformations between different exponential fields, 
corresponding to the same primary field of chiral algebra.
In this paper, we apply the method developed in \cite{Teschner} 
to derive the reflection amplitudes of the $N=2$ SLFT.
For simplicity, we will restrict ourselves to the case
of $\alpha={\overline\alpha}$ in Eq.(\ref{primary})
where the $U(1)$ charge of the (NS) operators becomes $0$.
We will refer to this case as the `neutral' sector.
(From now on, we will suppress the second indices ${\overline\alpha}$.)
The physical states in this sector are given by 
\beq
\alpha={1\over{2b}}+iP
\label{momentum}
\eeq
where $P$ is a real parameter.
This parameter is transformed by $P\to -P$ under
the reflection relation (\ref{reflection})
and can be thought of as a `momentum' which is reflected off from
a potential wall.

Two-point functions of the same operators can be expressed as
\begin{eqnarray}
\langle N_{\alpha}(z,{\overline z})N_{\alpha}(0,0)\rangle
&=&{D^{N}(\alpha)\over{|z|^{4\Delta^N_{\alpha}}}}\\
\langle R^{+}_{\alpha}(z,{\overline z}) R^{-}_{\alpha}(0,0)\rangle 
&=&{D^{R}(\alpha)\over{|z|^{4\Delta^R_{\alpha}}}}
\end{eqnarray}
where $\Delta^N_{\alpha},\Delta^R_{\alpha}$ are given by Eq.(\ref{delta}).
The reflection amplitudes are given by the normalization factors 
$D^{N}(\alpha), D^{R}(\alpha)$ and should satisfy
\beq
D^{N}(\alpha)D^{N}(1/b-\alpha)=1,\qquad
D^{R}(\alpha)D^{R}(1/b-\alpha)=1.
\eeq
To find these amplitudes explicitly, we consider the operator
product expansions (OPEs) with degenerate operators.

The (NS) and the (R) degenerate operators in the neutral sector are 
$N_{\alpha_{nm}}$ and $R^{\pm}_{\alpha_{nm}}$ with integers $n,m$ and
\begin{equation}
\alpha_{nm}={1-n\over{2b}}-{mb\over{2}},\qquad n,m\ge 0.
\end{equation}
The OPE of a (NS) field with a degenerate operator $N_{-b/2}$ is 
simply given by
\begin{equation}
N_{\alpha}N_{-b/2}=N_{\alpha-b/2}+ C_{-}^{N}(\alpha)N_{\alpha+b/2}.
\label{opei}
\end{equation}
Here the structure constant can be obtained from the screening integral
as follows:
\begin{equation}
C_{-}^{N}(\alpha)= \kappa_1 
\ga(1-\alpha b)\ga(1/2-\alpha b-b^2/2)
\ga(-1/2+\alpha b)\ga(\alpha b+b^2/2)\,,
\end{equation}
where 
\bea
\kappa_1 ={\mu^2 b^4\pi^2\over{2}}
\ga(-b^2-1)  \ga\left(1+{b^2\over{2}}\right) 
\ga\left({b^2\over{2}}+{3\over{2}}\right)
\eea
with $\ga(x)=\Ga(x)/\Ga(1-x)$ as usual.

To use this OPE, we consider a three-point function
$\langle N_{\alpha+b/2}N_{\alpha}N_{-b/2}\rangle$
and take the OPE by $N_{-b/2}$ either on $N_{\alpha+b/2}$ or
on $N_{\alpha}$ using Eq.(\ref{opei}).
This leads to a functional equation
\beq
C_{-}^{N}(\alpha)D^{N}(\alpha+b/2)=D^{N}(\alpha).
\label{relationi}
\eeq

This functional equation can determine the (NS) reflection amplitude 
in the following form:
\begin{equation}
D^{N}(\alpha)=(\frac{\kappa_1}{b^4})^{-2\al/b} \, 
\ga(2\al/b-1/b^2)
{\ga(b\al+1/2)\over{\ga(b\al)}}f(\al)
\label{eqone}
\end{equation}
with an aribitrary function $f(\al)$ satisfying $f(\al)=f(\al+b)$.
To fix this unknown function, we need an additional functional equation.
It is natural that this relation is provided by
the dual action ${\cal A}_{\rm II}(1/b)$.

For this purpose, we consider OPEs with another degenerate operator, namely,
\begin{eqnarray}
N_{\alpha}R^{+}_{-1/2b}&=&R^{+}_{\alpha-1/2b}+
{\tilde C}_{-}^{N}(\alpha)R^{+}_{\alpha+1/2b}
\label{opeii}\\
R^{-}_{\alpha}R^{+}_{-1/2b}&=&N_{\alpha-1/2b}+
{\tilde C}_{-}^{R}(\alpha)N_{\alpha+1/2b}.  
\label{opeiii}
\end{eqnarray}
The structure constants can be computed by the screening integrals 
using the dual action ${\cal A}_{\rm II}(1/b)$, which is equivalent to
${\cal A}_{\rm I}(b)$. 
The results are
\begin{eqnarray}
{\tilde C}_{-}^{N}(\alpha)&=&
\kappa_{2}(b){\gamma(2\al/b-1/b^2)\over{\gamma(2\al/b)}},\\
{\tilde C}_{-}^{R}(\alpha)&=&
\kappa_{2}(b){\gamma(2\al/b-1/b^2+1)\over{\gamma(2\al/b+1)}}\,
\end{eqnarray}
where 
\beq
\kappa_{2}(b)= \tilde \mu \pi \ga\left(\frac1{b^2}+1\right)\,.\quad
\eeq
These results are consistent with the $N=2$ superminimal CFT results
\cite{Marian}.

Now we consider three-point functions,
$\langle R^{-}_{\alpha+1/2b}N_{\alpha}R^{+}_{-1/2b}\rangle$
and 
$\langle N_{\alpha+1/2b}R^{-}_{\alpha}R^{+}_{-1/2b}\rangle$.
Taking OPE with $R^{+}_{-1/2b}$ on one of two other operators in the 
correlation functions and using the OPE relations (\ref{opeii})
and (\ref{opeiii}),
we obtain an independent set of functional relations as follows:
\begin{eqnarray}
\label{relationiii}
{\tilde C}_{-}^{N}(\al)D^{R}(\al+1/2b)&=&D^{N}(\alpha),\\
{\tilde C}_{-}^{R}(\al)D^{N}(\al+1/2b)&=&D^{R}(\alpha).
\end{eqnarray}
Solving for the $D^{N}(\alpha)$, we find a most 
general solution of Eqs.(\ref{relationiii}) is
\begin{equation}
D^{N}(\alpha)=\kappa_2^{-2\al b}{\Ga^2(\al b+1/2)\over{\Ga^2(\al b)}}
\ga(2\al/b-1/b^2)g(\al)
\label{eqtwo}
\end{equation}
where $g(\al)$ is another arbitrary function satisfying $g(\al)=g(\al+1/b)$.
Combining Eqs.(\ref{eqone}) and (\ref{eqtwo}), 
and requiring the normalization $D^N (\alpha = \frac1{2b} ) =1 $, 
we can determine the (NS) reflection amplitude completely as follows:
\begin{equation}
D^{N}(\alpha)=-\frac 2 {b^2}\, \kappa_2^{-2\al b+1}\, 
\ga\left({2\al\over{b}}-{1\over{b^2}}\right)
\ga\left(\al b+{1\over{2}}\right)\ga(1-\al b)\,,
\label{eqthree}
\end{equation}
where two parameters in the actions, $\mu$ and $\tilde \mu$,
are related by
\beq
\left({\kappa_1\over{b^4}}\right)^{1/b}=\kappa_2^b.
\eeq
The (R) reflection amplitude can be obtained by Eq.(\ref{relationiii}):
\begin{equation}
D^{R}(\alpha)=-\frac {b^2} 2\kappa_2^{-2\al b+1}\, 
\ga\left({2\al\over{b}}-{1\over{b^2}}+1\right)
\ga\left(-\al b+{1\over{2}}\right) \ga(\al b).
\label{eqfour}
\end{equation}
We can rewrite these amplitudes using the momentum $P$ defined in 
Eq.(\ref{momentum}):
\begin{eqnarray}
D^{N}(P)&=&\kappa_2^{-2iPb}
{\Ga\left(1+{2iP\over{b}}\right)\over{\Ga\left(1-{2iP\over{b}}\right)}} 
{\Ga\left(1+iPb\right)\over{\Ga\left(1-iPb\right)}} 
{\Ga\left({1\over{2}}-iPb\right)\over{\Ga\left({1\over{2}}+iPb\right)}},\\ 
\label{eqfive}
D^{R}(P)&=&\kappa_2^{-2iPb}
{\Ga\left(1+{2iP\over{b}}\right)\over{\Ga\left(1-{2iP\over{b}}\right)}} 
{\Ga\left(1-iPb\right)\over{\Ga\left(1+iPb\right)}} 
{\Ga\left({1\over{2}}+iPb\right)\over{\Ga\left({1\over{2}}-iPb\right)}}. 
\end{eqnarray}

\section{Consistency Check}

To justify the reflection amplitudes derived in the previous section 
based on the conjectured action ${\cal A}_{\rm II}$, we provide 
several consistency checks.

It has been noticed that an integrable model with
two parameters proposed in \cite{Fateev} 
can have $N=2$ supersymmetry if one of the parameters take a special value
\cite{BasFat}. 
This means that one can compute the reflection amplitudes of the
$N=2$ SLFT independently as a special case of those in \cite{BasFat}.
Indeed, we have confirmed that the two results agree exactly.

Furthermore, one can check the reflection amplitude for specicific values 
of $\alpha$ directly from the action. 
When $\alpha = \frac 1{2b} - \frac b2$, the Coulomb integral using the 
action ${\cal A}_{\rm I}(b)$ gives
\beaq
\langle N_{\al} (0) N_{\al} (1)  \rangle 
&=& (\frac {\mu b^2}{2})^2 
\int d^2 z_1 d^2 z_2  \langle e^{\al (\varphi + \varphi^\dagger) }(0) \,
 e^ {\al (\varphi + \varphi^\dagger) }(1) \,
\psi^{\dagger} {\overline\psi}^{\dagger} e^{b\varphi(z_1,\bar z_1)} \,
\psi {\overline \psi} \, e^{b\varphi^{\dagger}(z_2, \bar z_2)}\rangle
\nonumber\\
&=& \frac {\mu^2  b^3 \pi^2}{(\al - \frac{1-b^2}{2b} )} 
\ga(\frac{1+b^2}2) \frac {\ga(-1-b^2)}{\ga(-1/2 - b^2/2)} \,.
\eeaq
This result agrees with Eq.(\ref{eqthree}) for 
$\al \to  \frac 1{2b} - \frac b2$.
Similarly, when $\al\to 0$, one can compute the two-point function
directly from the action ${\cal A}_{\rm II}(1/b)$ and can get
\beaq
\langle N_{\al} (0) N_{\al} (1)  \rangle 
&=& -\frac { \tilde \mu }{b^2} 
\int d^2 z\langle e^{\al (\varphi + \varphi^{\dagger}) }(0) \,
 e^ {\al (\varphi + \varphi^{\dagger}) }(1) \,
e^{(\varphi + \varphi^\dagger)/b } 
\partial (\varphi + \varphi^\dagger) 
\bar\partial (\varphi + \varphi^\dagger)(z,\bar z)\rangle 
\nonumber\\
&=&  -\frac{\tilde \mu}{b^2} \pi \alpha b  = 0\qquad
{\rm as}\quad \alpha\to 0  \,.
\eeaq
Here, the insertion we have considered is the only term
in the action ${\cal A}_{\rm II}(1/b)$ which can give nonvanishing
contribution.
Again, this result agrees with Eq.(\ref{eqthree}) for $\alpha =0$.

In the semiclassical limit $b\to 0$, the reflection amplitudes
can be interpreted as the quantum mechanical reflection 
amplitudes of the wave function of the zero-modes from the
exponential potential wall arising from the action (\ref{actionone}).
One can easily find that the reflection amplitude corresponding to
the (NS) operator is given by
\begin{equation}
R^{N}(P)\sim {\Ga(1+2iP/b)\over{\Ga(1-2iP/b)}},
\end{equation}
which is consistent with Eq.(\ref{eqfive}) in the limit $b\to 0$. 

In summary, we have conjectured a strong coupling effective action 
which is dual to the $N=2$ SLFT.
Based on this conjecture, we have computed the reflection amplitudes
(the two-point functions) of the (NS) and the (R) primary fields exactly. 
We have fixed the relation between the two parameters $\mu$ and $\tilde\mu$.
Then, we have checked the validity of these amplitudes
by comparing with an independent result along with 
some other consistency checks.
It would be nice to provide more stringent check.
One possibility is to consider the $N=2$ supersymmetric sinh-Gordon 
(or sine-Gordon) model as a perturbed integrable model of the $N=2$ SLFT.
Being integrable, one can compare the finite-size correction of the central
charge either by the thermodynamic Bethe ansatz or by the quantization
conditions based on the conjecture reflection amplitudes \cite{ongoing}.

\section*{\bf Acknowledgement}

This work is supported in part by Korea Research Foundation 
2002-070-C00025, KOSEF R01-1999-00018 (CA,CR),
and by Eastern Europe exchange program 12-69-002
sponsored by KISTEP (MS).


\begin{thebibliography}{99}
\bibitem{LFT} A. Polyakov, Phys. Lett. {\bf B103} (1981) 207. 
\bibitem{CurTho}T. Curtright and C. Thorn, Phys. Rev. Lett. 
{\bf 48} (1982) 1309.
\bibitem{SeiWit} N. Seiberg and E. Witten, Nucl. Phys. {\bf B426} (1994) 19.
\bibitem{ORaf} L. O'Raifeartaigh and V. V. Sreedhar, Phys. Lett. {\bf B461}
(1999) 66. 
\bibitem{ZamZam} A. B. Zamolodchikov and Al. B Zamolodchikov, Nucl. Phys.
{\bf B477} (1996) 577.
\bibitem{RasSta} R. C. Rashkov and M. Stanishkov, Phys. Lett. {\bf B380} 
(1996) 49; R. H. Poghossian, Nucl. Phys. {\bf B496} (1997) 451.
\bibitem{n2slft} K. Hori and A. Kapustin, JHEP {\bf 0108} (2002) 045.
\bibitem{Marian} G. Mussardo, G. Sotkov, and M. Stanishkov, Int. J. Mod. Phys.
{\bf A4} (1989) 1135.
\bibitem{Teschner} J. Teschner, Phys. Lett. {\bf B363} (1995) 65. 
\bibitem{Fateev} V. Fateev, Nucl. Phys. {\bf B473} (1996) 509.
\bibitem{BasFat} P. Baseilhac and V. Fateev, Nucl. Phys. {\bf B532} (1998) 
567.
\bibitem{ongoing} C. Ahn, C. Kim, C. Rim and M. Stanishkov, in preparation.
\end{thebibliography}
\end{document}